\documentclass[a4paper,11pt]{article}
\usepackage{graphicx}
\usepackage{epstopdf}
\usepackage{amsfonts,amsmath,amssymb}
\usepackage{hyperref}
\usepackage{float}

\usepackage{epsfig,multicol,bbm}

\newcommand\fverb{\setbox\pippobox=\hbox\bgroup\verb}
\newcommand\fverbit{\egroup\item[\fbox{\unhbox\pippobox}]}

\newbox\pippobox

\begin{document}


\begin{centering}
	
	\vspace{2cm}
	
	\textbf{\Large{
	Correlation of Horizons and Black Hole Thermodynamics}}
	
	\vspace{0.8cm}
	
	{\large   Sanam Azarnia$^{a}$, S. Sedigheh Hashemi $^{a,b}$  }
	
	\vspace{0.5cm}
	
	\begin{minipage}{.9\textwidth}\small
		\begin{center}
			
			{\it  $^{a}$Department of Physics, 
				Shahid Beheshti University, 
				Tehran, Iran }\\
			
			$^{b}$School of Particles and Accelerators, Institute for Research in Fundamental Sciences (IPM)
			P.O. Box 19395-5531, Tehran, Iran
			
			\vspace{0.5cm}
			{\tt  sanam.azarnia@gmail.com, Hashemiphys@ipm.ir}
			\\ 
			
		\end{center}
	\end{minipage}

\begin{abstract}
	  For black holes with more than one horizon, the existence of a global temperature is not fully understood. This can affect the thermodynamics of these black holes to be unclear. In this paper, using the method of quantum tunneling, we find a global temperature for asymptotically AdS multi-horizon black holes appeared in three dimensional quasitopological gravities. We also find a global temperature for the asymptotically warped AdS$_3$ black holes. In all of the cases the global temperature differs from the conventional Hawking temperature.
\end{abstract}
	\end{centering}
\newpage

\tableofcontents
\section{Introduction}	

	It has been a well established fact for decades that black holes as thermal objects  have thermodynamic characteristics\cite{1,beck}. In fact by Beckenstein-Hawking's work,  the area of black hole event horizon is related to its entropy. Also by Hawking's hypothesis on black hole radiation and defining the Hawking temperature (which is  proportional to surface gravity of the black hole), physicists were challenged to figure out the source of this radiation. One of the methods people have used to study this problem is to think of Hawking radiation as a process of tunneling from one side of horizon to the other side in the semiclassical picture \cite{pg,revpg}.
	
	In order to find the Hawking temperature for a horizon through the tunneling picture one can  write the metric in Painlev{\'e}-Gullstrand coordinates \cite{19,20} (to get rid of the singularity at the horizon), then by deriving the energy spectrum of the massless particles, the tunneling trajectory is achieved. Finally,  from the  exponential of the imaginary part of the action along the trajectory one can read the temperature.
	
	Although everything is fairly straightforward in a single horizon spacetime, this is not the case for multi-horizon black holes,  and the thermodynamics of such spacetimes seems to be a little vague. Thus, one can ask about the possibility of the existence of a global temperature and an overall entropy in multi-horizon spacetimes. Recently a couple of work has been done elaborating this subject. It was discussed in \cite{volovik1,Volovik2}, that the correlation between inner and outer horizons may affect the temperature and the entropy of the spacetime   in the case of Reissner–Nordstr\"{o}m and Kerr black holes. The results, expectedly, deviated from the conventional Hawking temperature related  only to the outer horizon, and the  difference is attributed to the possible correlation of two horizons. Same argument applies to the global entropy as it is not naively the sum of entropies of all the  horizons. Following this work, the temperature and entropy of other multi-horizon spacetimes were investigated in \cite{singha}, more specifically the temperature and entropy of Schwarzschild-de Sitter, Reissner-Nordstr\"{o}m-de Sitter and the rotating BTZ spacetimes have been derived and again not in accordance with the Hawking temperature of the outer horizon, possibly due to the claimed correlation of two or more horizons with each other.
	The results suggest a certain pattern to deduce the global temperature and entropy as a function of Hawking temperature and the entropy of all the horizons.
	
	In the present paper we will consider several multi-horizon spacetimes, and we will use the tunneling picture towards understanding their radiations. 
	The first model is analytical regular black holes in three dimensions. The solutions have a real scalar field $\phi$, which is related to the Maxwell field strength through a duality transformation. The lagrangian for constructing the actions are of the form $\mathcal{L}[g^{ab},R_{ab},\partial_a \phi]$, with a condition of $g_{tt}g_{rr}=-1$, and considering a simple magnetic ansatz for $\phi$, which is proportional to angular coordinate. Therefore a broad class of theories named as "Electromagnetic Quasitopological" (EM-QT) gravities were found \cite{warp}.
The next solution that we will investigate its thermodynamic properties is the warped AdS$_3$ black hole. Three dimensional topologically massive gravity with a negative cosmological constant $-L^2$, and positive Newton constant $G$, for any value of the graviton mass $\mu$, contains an AdS$_3$ vacuum solution. It was shown that for $\mu L \neq 3$, there are vacuum solutions given by warped AdS$_3$, with a timelike or spacelike $U(1)$ isometry \cite{ads}.
	Our results are in agreement with the above mentioned previous researches in finding a global temperature and the entropy. We also show that the total entropy
	for these spacetimes are obtained by the correlations between the horizons of these black holes.
	
	The structure of the letter is as follows. In section \ref{1}, we will investigate the Electromagnetic quasitopological gravities in three dimensions, and we will consider some black hole solutions from the mentioned category of solutions \cite{cano2}. We will compute the
	thermal radiation by using the semiclassical tunneling approach for a black hole with three and four horizons. 
	Moreover, we  compute the entropy  by an approach based on the method of the singular
	coordinate transformations for these black holes. 
	 In section \ref{4} we focus on the warped AdS$_3$ black holes. In the last section we will draw our conclusions.

\section{Global temperature for black holes in EQG
	}\label{1}
\subsection{Electromagnetic Quasitopological Gravities in three dimensions}
Motivated by the higher dimensional counterparts \cite{cano2,cano1}, the Electromagnetic Generalized Quasitopological gravities (EMG-QT) in three dimensions, are defined with the condition that  a general Lagrangian of the form $\sqrt{|g|}\mathcal{L}[g^{ab},R_{ab},\partial_a \phi]$ can be written in a total derivative, when evaluated on the ansatz \cite{pols}-\cite{Hervik}
\begin{equation}\label{eq1}
	{\rm d}s^2=-f(r){\rm d}t^2+f(r)^{-1}{\rm d}r^2+r^2{\rm d}\varphi^2  ,\quad \phi=p \varphi\, ,
\end{equation}
where $p$ is an arbitrary dimensionless constant, and $\varphi$ is a real scalar. Theories, which satisfy such a property, yield solutions of the form written in Eq. (\ref{eq1}), in which the equation for $f(r)$ can be integrated and results into a differential equation of order $2$ (at most). In most of the cases, the equation for $f(r)$ is algebraic. These kind of theories are called "Electromagnetic Quasitopological"(EM-QT). The following family of densities, which belongs to the three dimensional EM-QT class \cite{warp} are given by
	\begin{equation}\label{EQTG}
	I_{\rm  EMQT}=\frac{1}{16\pi G}\int d ^3x\sqrt{|g|}\left[R+\frac{2}{L^2}- \mathcal{Q} \right]\, ,
\end{equation}
where
\begin{align} \notag
	\mathcal{Q}  \equiv & \sum_{n=1} \alpha_n L^{2(n-1)} (\partial \phi)^{2n}-\sum_{m=0} \beta_m L^{2(m+1)}(\partial \phi )^{2m}      \left[ (3+2m) R^{bc} \partial_b \phi \partial_c \phi- (\partial \phi )^2 R \right] \, ,
\end{align} 
in which, $\alpha_{n}$, $\beta_m$ are arbitrary constants with no dimension, and $L$ is related to the cosmological constant. It can be seen that
 $	\mathcal{Q} $ contains terms, which are at most linear in Ricci curvatures. By evaluating the non-linear equations of (\ref{EQTG}) for the ansatz given by (\ref{eq1}), a single independent equation for the $f(r)$ function can be obtained. The result is \cite{cano2}
\begin{equation}\label{fgen}
	f(r)= \left[\displaystyle \frac{r^2}{L^2}-\mu-\alpha_1 p^2 \log (r/L)+ \sum_{n=2} \frac{\alpha_n  p^{2n} L^{2(n-1)}}{2(n-1)r^{2(n-1)}} \right] \cdot \left[ \displaystyle 1+ \sum_{m=0} \frac{ \beta_m p^{2(m+1)} (2m+1) L^{2(m+1)}}{r^{2(m+1)}} \right]^{-1}\, ,
\end{equation}
which is the only static and spherically symmetric solution of equation (\ref{EQTG}). In the $f(r)$ metric function, $\mu$ is an integration constant, related to the mass of the solution via $M= \mu +\beta_0 p^2$+$\alpha_1 p^2 Log (r_0/L)$, and $r_0$ is a cutoff radius. By setting $\alpha_{n}$, and $\beta_m$ to zero, the static BTZ solution will be obtained. Moreover, if only $\alpha_1$
is active, the metric takes the same form as the charged BTZ black holes \cite{Clement}-\cite{ Martinez}.
In the next section, we will focus on some solutions of the above metric.

\subsection{Black hole Solutions and  Hawking radiation}\label{3}
By choosing different values of $\alpha_{n}$, and $\beta_m$, one can find black hole solutions, which we are going to consider, in order to study the Hawking radiation from their horizons. In the first case, we will consider a black hole solution with four horizons, and the next solution is a three horizon black hole.
\subsubsection{Case 1: Four-horizon black holes}
Setting  $\alpha_1=\alpha_2= \alpha_3=0$, keeping only the term $n=4$, and $\beta_m =0$ in equation (\ref{fgen}), we get 
	\begin{align}\label{3}
		f(r)=\frac{r^2}{L^2}-\mu +A \frac{L^6}{r^6},
	\end{align}
where we have defined $A= \alpha_4 p^8/6$. 
The number of horizons comes from the positive roots of $f(r)=0$, leading to four positive roots.

The temperature of the thermal Hawking radiation can be obtained by using the method of the semiclassical quantum tunneling. Since this solution has four horizons, the Hawking radiation will be modified \cite{14}-\cite{18}. In the semiclassical method the process of the Hawking radiation is studied by using the Painlev{\'e}-Gullstrand (PG) metric \cite{19,20}.
The PG metric is obtained in terms of the following coordinate transformation
\begin{equation}
	{\rm d}t \rightarrow {\rm d}t \pm a(r) {\rm d}r,
\end{equation}
where
\begin{eqnarray}\label{6}
	a(r)=\frac{\sqrt{1-f(r)}}{f(r)},
\end{eqnarray}
which leads to 
\begin{equation}\label{7}
	{\rm d}s^2= -{\rm d}t^2+({\rm d}r \pm v(r) {\rm d}t)^2+r^2{\rm d}\phi^2,
\end{equation}
where the shift velocity is 
\begin{equation}
	v(r)^2=1-f(r).
\end{equation}
The tunneling trajectory for the massless particle is 
\begin{equation}\label{99}
	g^{\alpha \beta}p_{\alpha}p_{\beta}=0~~\rightarrow E= p_{r}v(r)\pm p_{r}
\end{equation}
where $g^{\alpha \beta}$ is the contravriant metric, that is the inverse to the PG metric $g_{\alpha \beta}$ in equation (\ref{7}),
 $p_{r}$ is the radial momentum, and $E = p_0$.

The probability of the tunneling process is defined as the exponent of the imaginary part of the action along the tunneling trajectory $Im \int p_{r}(r, E){\rm d}r$, where the trajectory $p_r$ is (we choose the minus solution in equation (\ref{99}))
\begin{equation}
	p_r(r,E)= \frac{E}{v(r)-1}=-E\frac{\sqrt{1-f(r)}+1}{f(r)},
\end{equation}
therefore the probability of the hawking radiation becomes
\begin{equation}
\Gamma=\exp\left[	\text{Im} \int p_{r}(r,E){\rm d}r\right]=\exp\left[-\text{Im}\int E\frac{\sqrt{1-f(r)}+1}{f(r)}{\rm d}r\right].
\end{equation}
Since for this spacetime, $f(r)$ has four positive horizons, the imaginary part of the action is produced by all the four poles. The contribution of the four poles results into the following probability of the Hawking radiation as
\begin{equation}
	\Gamma = \exp \left[-4 \pi E \big(\frac{1}{f'(r)}|_{r=r_1}+\frac{1}{f'(r)}|_{r=r_2}+\frac{1}{f'(r)}|_{r=r_3}+\frac{1}{f'(r)}|_{r=r_4}\big)\right],
\end{equation}
where $r_1...r_4$ are the horizons $(r_1<...<r_4)$. Consequently, the above probability  can be written as
\begin{equation}\label{gamma}
	\Gamma= \exp\left[-\frac{2 \pi E}{\kappa_{eff}}\right],
\end{equation}
where 
\begin{equation}\label{kappa}
	\kappa_{eff} = \left(\frac{1}{\kappa_1}+\frac{1}{\kappa_2}+\frac{1}{\kappa_3}+\frac{1}{\kappa_4}\right)^{-1},
\end{equation}
in which $\kappa_i$, $(i=1..4)$ is the surface gravity at horizon $r=r_i$.
Equation (\ref{gamma}) is related to thermal radiation characterized by the Hawking temperature, where the Hawking temperature is defined as
\begin{equation}
	T_{H}= \frac{\kappa_{eff}}{2 \pi}.
\end{equation}
This equation shows that  for this spacetime a global temperature can exist due to the presence of the four horizons.

The probability of the Hawking radiation  (\ref{gamma}) can be written in the following form
\begin{equation}\label{16}
	\Gamma = \Gamma_1 \Gamma_2 \Gamma_3 \Gamma_4= \exp \left(-\frac{E}{T_1}\right)\exp \left(-\frac{E}{T_2}\right)\exp \left(-\frac{E}{T_3}\right)\exp \left(-\frac{E}{T_4}\right),
\end{equation}
where $T_1...T_4$ are the conventional  Hawking temperatures at the horizons (e.g., $T_1= \kappa_1/2 \pi$.) 
The temperature $T_4$ determines the rate of the tunneling from the region $r_3<r<r_4$, $r>r_4$. $T_3$ determines the rate of the tunneling from the region $r_2<r<r_3$, $r>r_3$. Moreover, $T_2$ specifies the rate of the tunneling from the region $r_1<r<r_2$, $r>r_2$, while $T_1$ shows the occupation number of the particles near the inner horizon. As a result,
the presence of the four horizons, and their correlation leads to the modified Hawking radiation as 
\begin{equation}
	\Gamma = \exp\left(-\frac{E}{T_{H}}\right).
\end{equation}
This equation  shows that the global temperature does not coincide with the conventional Hawking temperature related only to the outer horizon. 

The coordinate transformations written in equation (\ref{6}) have four positive singularities. Thus, the macroscopic quantum tunneling \cite{volovik1,Volovik2} from the PG metric of our black hole solution to its static partner, which has the same energy $E$, is given by
\begin{align}
	\Gamma_{BH\rightarrow static}=&\exp \left(-2 Im \int E {\rm d}\tilde{t}\right)\nonumber\\&
	=\exp \left(-2 E Im \int \left({\rm d}t+a(r){\rm d}r\right)\right)\nonumber\\&=\exp \left(-2 E Im \int a(r){\rm d}r\right)
	\nonumber\\&=\exp \left(-\frac{2\pi E}{\kappa_{eff}}\right).
\end{align}
The probability of the quantum tunneling between the PG metric of our black hole metric, and its static partner can be regarded as the quantum fluctuation \cite{landa}. Consequently, 
 the entropy of the  black hole is
\begin{equation}\label{19}
	S = \frac{2\pi}{\kappa_{eff}^2}=\left(\sqrt{S_1}+\sqrt{S_2}+\sqrt{S_3}+\sqrt{S_4}\right)^2,
\end{equation}
where we have assumed that the correspondence between entropy and surface gravity for single horizon spacetime can be true for multi-horizon spacetime \cite{volovik1,singha}, that is 
$
	S_1= 2\pi/\kappa_1{^2}
$.
Hence the total entropy for our metric is not determined only by the outer horizon, and the correlation of the four horizons determine it \cite{sh}.

It should be noted that we can  find another solution from equation (\ref{fgen}) given by
	\begin{align}\label{4}
	f(r)=\frac{\frac{r^2}{L^2}-\mu +A \frac{L^6}{r^6}}{1+\frac{B L^2}{r^2}},
\end{align}
where $B= \beta_{0} p^2$. This solution has four horizons, and the calculations of the probability of tunneling, and entropy will give the same result as (\ref{16}), and (\ref{19}).

\subsubsection{Case 2: Three-horizon black holes}The next solution that we will focus on it, is a solution  with three horizons
\begin{equation}
	f(r)=\frac{r^2}{L^2}-\mu +A \frac{L^4}{r^4},
\end{equation}
where $A = \alpha_3 p^6/4$.
The horizons of the above solution can be calculated via $f(r)=0$, which results into three positive horizons.
Therefore by doing the same procedure as previous section, the probability of the tunneling is
\begin{equation}
	\Gamma=\exp\left[\text{Im} \int p_{r}(r,E){\rm d}r\right]=\exp\left[-\text{Im}\int E\frac{\sqrt{1-f(r)}+1}{f(r)}{\rm d}r\right]=\exp\left[-2 \pi E \left(\frac{1}{\kappa_1}+\frac{1}{\kappa_2}+\frac{1}{\kappa_3}\right)\right],
\end{equation}
which can be written as
\begin{equation}
	\Gamma= \exp\left[-\frac{2 \pi E}{\kappa_{eff}}\right],
\end{equation}
where
\begin{equation}\label{23}
	\kappa_{eff}= \left(\frac{1}{\kappa_1}+\frac{1}{\kappa_2}++\frac{1}{\kappa_3}\right)^{-1}.
\end{equation}
Equation (\ref{23}) corresponds to thermal radiation, and Hawking temperature, given by
\begin{equation}
	T_{H}= \frac{\kappa_{eff}}{2 \pi}.
\end{equation}
Again this shows that a global temperature can exist due to the presence of the three horizons. For this spacetime the probability of the tunneling can be written as 
\begin{equation}
	\Gamma= \Gamma_{1}\Gamma_{2}\Gamma_{3}=\exp\left(-\frac{E}{T_1}\right)\exp\left(-\frac{E}{T_2}\right)\exp\left(-\frac{E}{T_3}\right),
\end{equation}
where $T_1..T_3$ are the conventional Hawking temperature at the horizons. 
The temperature $T_3$ determines the rate of the tunneling from the region $r_2<r<r_3$, $r>r_3$. $T_2$ determines the rate of the tunneling from the region $r_1<r<r_2$, $r>r_2$,
moreover $T_1$ determines the occupation number of these particles near the black hole horizon. Now, we can write the final probability of the Hawking radiation as
\begin{equation}
	\Gamma= \Gamma_{1}\Gamma_{2}\Gamma_{3}=\exp\left(-\frac{E}{T_H}\right),
\end{equation}
which suggests that the global temperature does not coincide with the Hawking temperature,  given by the outer horizon.

In order to find the entropy for this spacetime, we use the approach of single coordinate transformation as
\begin{align}
	\Gamma_{BH\rightarrow static}=&\exp \left(-2 ~\text{Im} \int E~ {\rm d}\tilde{t}\right)\nonumber\\&
	=\exp \left(-2 E ~\text{Im}\int \left({\rm d}t+a(r){\rm d}r\right)\right)\nonumber\\&=\exp \left(-2 E~ \text{Im} \int a(r){\rm d}r\right)
	\nonumber\\&=\exp \left(-2 E ~\text{Im}\int \frac{\sqrt{1-f(r)}}{f(r)}{\rm d}r\right)\nonumber\\&=\exp\left(-\frac{2 \pi E}{\kappa_{eff}}\right),
\end{align}
therefore the entropy is
\begin{equation}
	S= \frac{2 \pi}{\kappa_{eff}^2}=\left(\sqrt{S_1}+\sqrt{S_2}+\sqrt{S_3}\right)^2,
\end{equation}
where $S_1= 2 \pi/\kappa_1^2$,  $S_2= 2 \pi/\kappa_2^2$, and  $S_3= 2 \pi/\kappa_3^2$.
Consequently, the total entropy for this spacetime can not be determined only by the outer horizon, and the correlation of the horizons determine it. 

In the next section, we will study the probability of tunneling and the entropy for the warped AdS$_3$ spacetime.

\section{Warped AdS$_3$}\label{4}
In the previous sections we have calculated the temperature and the entropy of asymptotically AdS spacetimes. In this section we consider spacetimes, which are not same as previous section, but asymptotic to warped AdS$_3$.
These black holes are solutions of topologically massive gravity (TMG).
The action of a three dimensional TMG with a negative cosmological constant is 
 \begin{equation}
	I_{\text{TMG}}=\frac{1}{16 \pi G}\int \sqrt{-g}\left(R+\frac{2}{L^2}\right)+\frac{L}{96 \pi G \gamma}\int \sqrt{-g}\epsilon^{\lambda \mu \nu}\Gamma^{r}_{\lambda \sigma}\left(\partial_\mu \Gamma^{\sigma}_{r \nu}+\frac{2}{3}\Gamma ^{\sigma}_{\mu \tau}\Gamma^{\tau}_{\nu r}\right),
\end{equation}
where $\epsilon^{\tau \sigma u} = 1/\sqrt{-g}$ is the Levi-Civita tensor, and the dimensionless coupling $\gamma$ is related to mass and cosmological constant by $\gamma = \mu L/3$. By varying the action with respect to the metric, one can obtain the equations of motion. From these equations, any Einstein vacuum solution is a solution of TMG. There are also non-Einstein  solutions. One of the simplest of them is the warped AdS$_3$, as it contains a warped fibration \cite{ads}.

 The metric which describes the spacelike stretched black holes for $\gamma^2>1$ is given by

\begin{equation}\label{warp}
	{\rm d}s^2 = -N(r)^2{\rm d}t^2+\frac{L^4{\rm d}r^2}{4 R(r)^2 N(r)^2}+L^2R(r)^2\left({\rm d}\theta+P(r){\rm d}t\right)^2,
\end{equation}
where
\begin{equation}
	N(r)^2= \frac{L^2(r-r_+)(r-r_{-})(\gamma^2+3)}{4 R(r)^2},
\end{equation}
\begin{equation}
	R(r)^2= \frac{r}{4}\Big(3 r(\gamma^2-1)+(\gamma^2+3)(r_++r_-)- 4 \gamma\sqrt{r_+r_-(\gamma^2+3)}\Big),
\end{equation}
and
\begin{equation}
	P(r)= \frac{2\gamma r -\sqrt{r_+r_-(\gamma^2+3)}}{2 R(r)^2}.
\end{equation}
 For this solution, we want to calculate the probability of the tunneling, and the entropy.
The horizons are located at $r_+$, and $r_-$, where $1/g_{rr}$ vanishes.
In order to find the PG metric for this solution, 
for convenience we choose zero angular momentum $\theta= \Omega t$, thus the metric (\ref{warp}), becomes 
\begin{equation}
	{\rm d}s^2 = -N(r)^2{\rm d}t^2+\frac{L^4{\rm d}r^2}{4 R(r)^2 N(r)^2},
\end{equation}
we set 
\begin{equation}
	W(r) =\frac{4 R(r)^2 N(r)^2}{L^4}.
\end{equation}
In terms of the following coordinate transformation
\begin{equation}\label{40}
	{\rm d}t ={\rm d}\tau \pm\frac{1}{N(r)}\sqrt{\frac{1-W(r)}{W(r)}}{\rm d}r,
\end{equation}
we obtain the PG metric for the spacetime as
\begin{equation}
	{\rm d}s^2 = -N(r)^2{\rm d}\tau^2+{\rm d}r^2\pm \frac{2 N(r)}{\sqrt{W(r)}}\sqrt{1-W(r)}~{\rm d}r{\rm d}\tau.
\end{equation}
The tunneling trajectory for the massless particle is (by choosing the plus sign in the above equation)
\begin{equation}\label{9}
	g^{\alpha \beta}p_{\alpha}p_{\beta}=0~~\rightarrow E= \frac{N(r)}{\sqrt{W(r)}}\left(\pm 1+\sqrt{1-W(r)}\right)p_r,
\end{equation}
where $E=p_0$, and $p_r$ is the radial momentum. Consequently,
the  probability of the tunneling process is defined as the exponent of the imaginary part of the action along the tunneling trajectory $\text{Im} \int p_{r}(r, E){\rm d}r$, where the trajectory $p_r$ is 
\begin{equation}
	p_r(r,E)= \frac{E \sqrt{W(r)}}{N(r)\left(\pm 1+\sqrt{1-W(r)}\right)},
\end{equation}
therefore the probability of the hawking radiation becomes
\begin{equation}
	\Gamma=\exp\left[	\text{Im} \int p_{r}(r,E){\rm d}r\right]=\exp\left[\text{Im}\int  \frac{E \sqrt{W(r)}}{N(r)\left(\pm 1+\sqrt{1-W(r)}\right)}{\rm d}r\right],
\end{equation}
now by choosing the minus sign of $\pm$ in the denominator of the integrand, we get
\begin{equation}
\Gamma= \exp \left(-2 E \frac{1}{(N(r)\sqrt{W(r)})'}\right)=\exp\left(-4 \pi E X\right)= \exp\left(-2 \pi E (\frac{1}{\kappa_+}+\frac{1}{\kappa_-})\right)=\exp\left(-\frac{2 \pi E}{\kappa_{eff}}\right),
\end{equation}
where prime denotes derivative with respect to $r$, and $X$ is defined as
\begin{equation}
	X = \frac{\sqrt{r_+ \left(4 \gamma  \left(\gamma  r_+-\sqrt{\left(\gamma ^2+3\right) r_+ r_-}\right)+\left(\gamma ^2+3\right)
			r_-\right)}-\sqrt{r_- \left(\left(\gamma ^2+3\right) r_++4 \gamma \left(\gamma r_--\sqrt{\left(\gamma ^2+3\right) r_+
				r_-}\right)\right)}}{\left(\gamma ^2+3\right) (r_+-r_-)}.
\end{equation}
As a result, we can read the Hawking radiation, given by
\begin{equation}
	T_{H}= \frac{\kappa_{eff}}{2 \pi },
\end{equation}
hence for this spacetime, a global temperature can exist because of the presence of the two horizons.

The coordinate transformation in (\ref{40}) have two singularities, which are the horizons. Thus, the macroscopic quantum tunneling from the PG metric of this spacetime to its static partner with the same energy $E$ is given by the following exponent
\begin{equation}
\Gamma_{WAdS_3\rightarrow static}=\exp\left(-2 E~ \text{Im}\int \frac{\sqrt{1-W}}{N\sqrt{W}}{\rm d}r\right)= \exp\left(-\frac{2 \pi E}{\kappa_{eff}}\right),
\end{equation}
therefore the entropy for this spacetime is
\begin{equation}
	S = \frac{2 \pi}{\kappa_{eff}^2},
\end{equation}
which shows that the entropy is not specified only by the outer horizon, but it is determined by the correlation of the two horizons.
 It should be noted that,  metric (\ref{warp}) reduces to rotating  BTZ black hole when we set $\gamma^2 =1$. In this case our calculations for the Hawking temperature and entropy are in agreement with the results obtained in \cite{singha}. Moreover, the structure of the $\kappa_{eff}$ in this case is the same as asymptotically AdS black holes in the previous section. This shows that the asymptotic behavior of black holes is not important in the process of tunneling.

\section{Conclusion}
In this paper, we considered the thermodynamics of three-dimensional  multi-horizon black holes with three, and four horizons. We also investigated the thermodynamics of warped AdS$_3$ black hole with two horizons. For all the mentioned black holes we calculated the entropy and the corresponding temperature of the thermal Hawking radiation using the method of semiclassical tunneling. We found that a global temperature can be obtained due to the presence of multi-horizons. For all the discussed black holes, the global temperature does not agree with the conventional Hawking temperature, which is related to the outer horizon. Moreover, the total entropy for these black holes can not be specified by the outer horizon. We found that the correlations between the horizons determine it. For further investigations, it would be quite interesting to apply the quantum tunneling method for other spacetimes such as black rings \cite{blackring}, and black saturns \cite{sat} in five dimensions.

\vspace{1cm}
\textbf{Acknowledgments}
The authors would like to thank R. Fareghbal for useful discussions. \\S. S. Hashemi also thanks Iran National Science Foundation for their support during this project.

\end{document}